\documentclass[aps,prl,superscriptaddress,showpacs,twocolumn,floatfix]{revtex4}
\usepackage{graphicx}

\newcommand{\bork}{Bj\"{o}rken }

\def\deg{^\circ}
\def\gtorder{\mathrel{\raise.3ex\hbox{$>$}\mkern-14mu
 \lower0.6ex\hbox{$\sim$}}}
\def\ltorder{\mathrel{\raise.3ex\hbox{$<$}\mkern-14mu
 \lower0.6ex\hbox{$\sim$}}}

\begin{document}

\title{New measurements of the EMC effect in very light nuclei}

\author{J. Seely}
\affiliation{Laboratory for Nuclear Science, Massachusetts Institute of Technology, Cambridge, MA, USA}

\author{A. Daniel}
\affiliation{University of Houston, Houston, TX, USA}

\author{D. Gaskell}
\affiliation{Thomas Jefferson National Laboratory, Newport News, VA, USA}

\author{J. Arrington}
\thanks{Corresponding author: johna@anl.gov}
\affiliation{Physics Division, Argonne National Laboratory, Argonne, IL, USA}

\author{N. Fomin}
\affiliation{University of Virginia, Charlottesville, VA, USA}

\author{P. Solvignon}
\affiliation{Physics Division, Argonne National Laboratory, Argonne, IL, USA}

\author{R. Asaturyan}
\thanks{Deceased}
\affiliation{Yerevan Physics Institute, Armenia}

\author{F. Benmokhtar}
\affiliation{University of Maryland, College Park, MD, USA}

\author{W. Boeglin}
\affiliation{Florida International University, Miami, FL, USA}

\author{B. Boillat}
\affiliation{Basel University, Basel, Switzerland}

\author{P. Bosted}
\affiliation{Thomas Jefferson National Laboratory, Newport News, VA, USA}

\author{A. Bruell}
\affiliation{Thomas Jefferson National Laboratory, Newport News, VA, USA}

\author{M.H.S. Bukhari}
\affiliation{University of Houston, Houston, TX, USA}

\author{M.E. Christy}
\affiliation{Hampton University, Hampton, VA, USA}

\author{B. Clasie}
\affiliation{Laboratory for Nuclear Science, Massachusetts Institute of Technology, Cambridge, MA, USA}

\author{S. Connell}
\thanks{Present address: University of Johannesburg, Johannesburg, South Africa}
\noaffiliation

\author{M.M. Dalton}
\affiliation{University of Virginia, Charlottesville, VA, USA}

\author{D. Day}
\affiliation{University of Virginia, Charlottesville, VA, USA}

\author{J. Dunne}
\affiliation{Mississippi State University, Jackson, MS, USA}

\author{D. Dutta}
\affiliation{Mississippi State University, Jackson, MS, USA}
\affiliation{Triangle Universities Nuclear Laboratory, Duke University, Durham, NC, USA}

\author{L. El Fassi}
\affiliation{Physics Division, Argonne National Laboratory, Argonne, IL, USA}

\author{R. Ent}
\affiliation{Thomas Jefferson National Laboratory, Newport News, VA, USA}

\author{H. Fenker}
\affiliation{Thomas Jefferson National Laboratory, Newport News, VA, USA}

\author{B.W. Filippone}
\affiliation{Kellogg Radiation Laboratory, California Institute of Technology, Pasadena, CA, USA}

\author{H. Gao}
\affiliation{Laboratory for Nuclear Science, Massachusetts Institute of Technology, Cambridge, MA, USA}
\affiliation{Triangle Universities Nuclear Laboratory, Duke University, Durham, NC, USA}

\author{C. Hill}
\affiliation{University of Virginia, Charlottesville, VA, USA}

\author{R.J. Holt}
\affiliation{Physics Division, Argonne National Laboratory, Argonne, IL, USA}

\author{T. Horn}
\affiliation{University of Maryland, College Park, MD, USA}
\affiliation{Thomas Jefferson National Laboratory, Newport News, VA, USA}

\author{E. Hungerford}
\affiliation{University of Houston, Houston, TX, USA}

\author{M.K. Jones}
\affiliation{Thomas Jefferson National Laboratory, Newport News, VA, USA}

\author{J. Jourdan}
\affiliation{Basel University, Basel, Switzerland}

\author{N. Kalantarians}
\affiliation{University of Houston, Houston, TX, USA}

\author{C.E. Keppel}
\affiliation{Hampton University, Hampton, VA, USA}

\author{D. Kiselev}
\affiliation{Basel University, Basel, Switzerland}

\author{M. Kotulla}
\affiliation{Basel University, Basel, Switzerland}

\author{C. Lee}
\affiliation{University of the Witwatersrand, Johannesburg, South Africa}

\author{A.F. Lung}
\affiliation{Thomas Jefferson National Laboratory, Newport News, VA, USA}

\author{S. Malace}
\affiliation{Hampton University, Hampton, VA, USA}

\author{D.G. Meekins}
\affiliation{Thomas Jefferson National Laboratory, Newport News, VA, USA}

\author{T. Mertens}
\affiliation{Basel University, Basel, Switzerland}

\author{H. Mkrtchyan}
\affiliation{Yerevan Physics Institute, Armenia}

\author{T. Navasardyan}
\affiliation{Yerevan Physics Institute, Armenia}

\author{G. Niculescu}
\affiliation{James Madison University, Harrisonburg, VA, USA}

\author{I. Niculescu}
\affiliation{James Madison University, Harrisonburg, VA, USA}

\author{H. Nomura}
\affiliation{Tohoku University, Sendai, Japan}

\author{Y. Okayasu}
\affiliation{University of Houston, Houston, TX, USA}
\affiliation{Tohoku University, Sendai, Japan}

\author{A.K. Opper}
\affiliation{Ohio University, Athens, OH, USA}

\author{C. Perdrisat}
\affiliation{College of William and Mary, Williamsburg, VA, USA}

\author{D.H. Potterveld}
\affiliation{Physics Division, Argonne National Laboratory, Argonne, IL, USA}

\author{V. Punjabi}
\affiliation{Norfolk State University, Norfolk, VA, USA}

\author{X. Qian}
\affiliation{Triangle Universities Nuclear Laboratory, Duke University, Durham, NC, USA}

\author{P.E. Reimer}
\affiliation{Physics Division, Argonne National Laboratory, Argonne, IL, USA}

\author{J. Roche}
\affiliation{Thomas Jefferson National Laboratory, Newport News, VA, USA}

\author{V.M. Rodriguez}
\affiliation{University of Houston, Houston, TX, USA}

\author{O. Rondon}
\affiliation{University of Virginia, Charlottesville, VA, USA}

\author{E. Schulte}
\affiliation{Physics Division, Argonne National Laboratory, Argonne, IL, USA}

\author{E. Segbefia}
\affiliation{Hampton University, Hampton, VA, USA}

\author{K. Slifer}
\affiliation{University of Virginia, Charlottesville, VA, USA}

\author{G.R. Smith}
\affiliation{Thomas Jefferson National Laboratory, Newport News, VA, USA}

\author{V. Tadevosyan}
\affiliation{Yerevan Physics Institute, Armenia}

\author{S. Tajima}
\affiliation{University of Virginia, Charlottesville, VA, USA}

\author{L. Tang}
\affiliation{Hampton University, Hampton, VA, USA}

\author{G. Testa}
\affiliation{Basel University, Basel, Switzerland}

\author{R. Trojer}
\affiliation{Basel University, Basel, Switzerland}

\author{V. Tvaskis}
\affiliation{Hampton University, Hampton, VA, USA}

\author{W.F. Vulcan}
\affiliation{Thomas Jefferson National Laboratory, Newport News, VA, USA}

\author{F.R. Wesselmann}
\affiliation{University of Virginia, Charlottesville, VA, USA}

\author{S.A. Wood}
\affiliation{Thomas Jefferson National Laboratory, Newport News, VA, USA}

\author{J. Wright}
\affiliation{University of Virginia, Charlottesville, VA, USA}

\author{L. Yuan}
\affiliation{Hampton University, Hampton, VA, USA}

\author{X. Zheng}
\affiliation{Physics Division, Argonne National Laboratory, Argonne, IL, USA}

\date{\today}

\begin{abstract}

New Jefferson Lab data are presented on the nuclear dependence of the
inclusive cross section from $^2$H, $^3$He, $^4$He, $^9$Be and $^{12}$C
for $0.3 < x < 0.9$, $Q^2 \approx 3$--6~GeV$^2$. These data represent the first
measurement of the EMC effect for $^3$He at large $x$ and a significant
improvement for $^4$He.  The data do not support previous A-dependent or
density-dependent fits to the EMC effect and suggest that the nuclear
dependence of the quark distributions may depend on the \textit{local} nuclear
environment.

\end{abstract}


%13.60.-r   Photon and charged-lepton interactions with hadrons (for neutrino interactions, see 13.15) 
%24.85.+p   Quarks, gluons, and QCD in nuclei and nuclear processes
%25.30.Fj   Inelastic electron scattering to continuum

\maketitle

High energy lepton scattering provides a clean method of probing the
quark momentum distributions in nucleons and nuclei.  The early expectation
was that probes at the GeV energy scale would be insensitive to nuclear
binding effects, which are typically on the order of several MeV.  The
effects were expected to be small except at large Bj\"{o}rken-$x$, corresponding to
very high momentum quarks.  In this region, the rapid falloff of the
parton distributions approaching the kinematical limit of $x \to 1$ makes
the distributions very sensitive to the smearing effect of the nucleon's motion.

In 1983 the European Muon Collaboration (EMC) discovered that the per-nucleon
deep inelastic structure function, $F_2(x)$, in iron was significantly different
than that for deuterium~\cite{aubert83}.  They showed a clear suppression
of high momentum quarks for $0.3<x<0.8$, confirmed for several nuclei in more
extensive measurements at SLAC~\cite{gomez94}. This phenomenon, dubbed the
``EMC effect'', has become the subject of a determined theoretical effort aimed
at understanding the underlying physics.  While progress has been made in
explaining the principal features of the effect, no single model has been able
to explain the effect over all $x$ and A~\cite{geesaman95,norton03}.  Much of
the effort has focused on heavy nuclei, and many models are evaluated for 
infinite nuclear matter and scaled to the density of finite nuclei, neglecting
possible surface effects or the impact of detailed nuclear structure.

There has been less focus on few-body nuclei, which provide the opportunity to
test models in cases where the details of the nuclear structure are well
understood. These data are also necessary to get a complete picture of the
evolution of nuclei from deuterium to infinite nuclear matter.  Precise
measurements in few-body nuclei allow for stringent tests of calculations
of the effects of Fermi motion and nuclear binding, which is the dominant
effect at large $x$. In addition, these data allow us to test simple scaling
models of the EMC effect.
A global analysis of the SLAC data~\cite{gomez94} found that the data could be
equally well described by fits that assumed the EMC effect to be proportional
to the average nuclear density, $\rho$, or by fits that assumed it scaled with
the nuclear mass, i.e. an EMC effect proportional to ln(A).  These simple fits
for the nuclear dependence did equally well for heavy nuclei (A~$\gtorder 12$),
where the density varies slowly with A.  For very light nuclei, these simple
models predict different behavior, but the limited data on light nuclei were
not sufficient to differentiate between these predictions.

To address these issues, Jefferson Lab (JLab) experiment E03-103 was proposed
to make high precision measurements of the EMC effect at large $x$ in both
heavy and few-body nuclei.  The experiment ran in Hall C during the fall of 
2004.
The measurement used a 5.767 GeV, 80~$\mu$A unpolarized electron beam, with
scattered electrons detected in the High Momentum Spectrometer (HMS).  The
primary measurements were taken at a scattering angle of 40$\deg$,
with additional data taken at different angles and/or 5 GeV beam energy to
examine the $Q^2$ dependence. Data were collected on four cryotargets - $^1$H,
$^2$H, $^3$He, and $^4$He, and solid Beryllium, Carbon, Copper, and Gold
targets arranged together on a common target ladder.  The target ladder held
only two cryotargets at a time, so there were two separate running periods to
collect data on all four cryogenic targets.  Data were taken on solid targets
during both periods for systematic checks on the relative normalization during
the two run periods.  In this paper, we focus on the light nuclei, $A\leq 12$,
for which fewer data exist and which require smaller corrections due to
backgrounds and Coulomb distortion.

The HMS subtends a solid angle of 7~msr and the momentum bite was restricted
to the central part of the acceptance ($\pm$9\%).  The detector package
consisted of two sets of wire chambers for tracking, four planes of hodoscopes
for triggering, and a gas \v{C}erenkov and lead-glass calorimeter for online
and offline particle identification~\cite{dutta03}.  The cross sections were
corrected for electronic and computer deadtimes, detector efficiencies, and
radiative effects (which closely followed the approach of Ref.~\cite{dasu94}).
Data were taken at several beam currents on carbon to look for rate-dependent
corrections, and on all four cryotargets to measure current-dependent target
density effects due to heating at high current.

The dominant sources of background were pion production, electrons scattering
from the aluminum cryocell wall and electrons from pair-production in the
target.  After applying calorimeter and \v{C}erenkov cuts, the pion
contamination was negligible for the kinematics shown here.  The electron
background (8--19\%) from the cell wall was subtracted using measurements on
a ``dummy'' target, consisting of two Aluminum targets at the positions of the
cryocell walls, with radiative corrections calculated
separately for the real cryocells and the dummy target. The background from pair
production was measured by reversing the HMS polarity to detect
positrons, yielding a direct measure of the charge-symmetric background,
strongly dominated by pair production.  This background was typically 5--10\%,
but was as much as 30\% of the total yield at the lowest $x$ and largest
$Q^2$ values.

There are several sources of systematic uncertainty which we separate into
point-to-point and normalization uncertainties.  Normalization
uncertainties are those that modify the overall scale, but not the $x$ or
$Q^2$ dependence of the target cross section ratios, \textit{e.g.}, target
thicknesses. Point-to-point uncertainties can vary with $x$ or $Q^2$, and are
treated in the same way as statistical uncertainties.

The cryogenic target thicknesses were determined from the dimensions of the
cryocell and the density of the cryogen, as computed from measurements of its
pressure and temperature. The total normalization uncertainty in the
cross section ratios was between 1.6 and 1.9\%, mainly
due to the 1-1.5\% uncertainty in the target thicknesses.  Uncertainty
in the target boiling correction contributes $\sim$0.4\%, radiative
corrections~\cite{dasu94} contribute 0.1--0.75\%, depending on the kinematics
and target thickness, and the acceptance contributes 0.5\% (0.2\%) to
the solid target (cryotarget) ratios.  The dominant sources
of point-to-point uncertainties come from charge measurement drifts (0.3\%),
corrections due to drift of beam position on target (0.45\%), radiative
corrections (0.5\%), deadtime determination (0.3\%), detector efficiencies
(0.3\%), acceptance (0.3\%).  Charge-symmetric background subtraction
contributes 0.1-0.6\% to the uncertainty, and is largest for the Be and C
targets.  The uncertainties in the
kinematics contribute up to 0.6\% to the uncertainties in the ratios, with
larger effects at large $x$ values where the cross section is changing most
rapidly.  We apply Coulomb distortion corrections following the effective
momentum approximation of Aste~\cite{aste05}.  The corrections are
$\ltorder$1\% for $^{12}$C, and much smaller for the helium data.

The results are shown as ratios of the cross section per nucleon, rather than
the $F_2$ structure functions.  These ratios are identical if the ratio of
longitudinal to transverse cross sections, $R = \sigma_L/\sigma_T$, is
independent of A. If $R_A \ne R_D$, then there will be a correction involved
in going from cross section ratios to the $F_2$ ratios~\cite{geesaman95}.

In the \bork limit, the structure function exhibits scaling, i.e. becomes
independent of $Q^2$ except for the weak $Q^2$ dependence from QCD evolution of
the parton distributions.  This scaling has been observed in the deep-inelastic
scattering (DIS) region, which for e--p scattering is typically taken to be
$Q^2>1$~GeV$^2$ and $W^2>4$~GeV$^2$, where $W$ is the invariant mass of the
unmeasured system.  In nuclei, it has been observed that results are nearly
independent of $Q^2$ to lower values of $W^2$ for $Q^2 \gtorder
3$~GeV$^2$~\cite{arrington01}.  A precise measurement of the target ratios in
the resonance region~\cite{arrington06a} for $Q^2=3$--4~GeV$^2$ showed that
the nuclear dependence is identical to the high $Q^2$ measurements up to
$x\approx 0.8$, even though the DIS region is limited to $x<0.5$ for these
$Q^2$ values.

\begin{figure}[tbh]
\centering
\includegraphics[width=85mm]{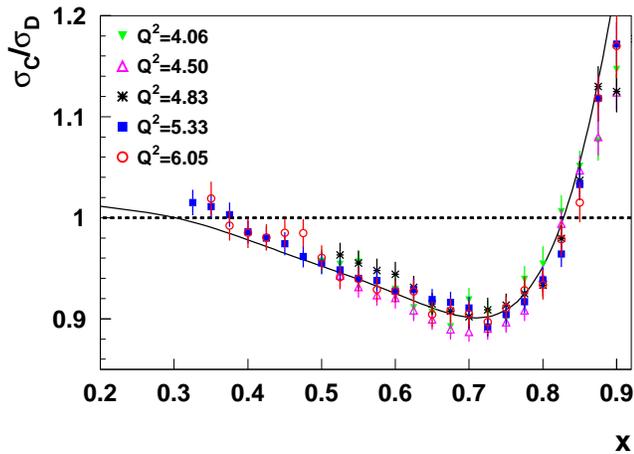}
\caption{(Color online) Carbon EMC ratios~\cite{note_webdata} for the five
highest $Q^2$ settings ($Q^2$ quoted at $x=0.75$). Uncertainties are
the combined statistical and point-to-point systematic.  The solid curve is
the SLAC fit~\cite{gomez94} to the Carbon EMC ratio.}
\label{fig:qsqdep}
\end{figure}

Because these data are at somewhat lower $Q^2$ than previous high-$x$ results,
typically $Q^2$=5 or 10~GeV$^2$ for SLAC E139~\cite{gomez94}, extensive
measurements were made to verify that our result is independent of $Q^2$.  The
structure functions were extracted at several $Q^2$ values and found to be
consistent with scaling violations expected from QCD down to $Q^2 \approx
3$~GeV$^2$ for $W^2 \geq 1.5$~GeV$^2$, while the structure functions ratios
show no $Q^2$ dependence.  Figure~\ref{fig:qsqdep} shows the carbon to
deuteron ratio for the five highest $Q^2$ settings (the lowest and middle
$Q^2$ values were measured with a 5~GeV beam energy).  There is no systematic
$Q^2$ dependence in the EMC ratios, even at the largest $x$ values, consistent
with the observation of previous measurements~\cite{geesaman95}.

For all further results, we show the ratios obtained from the 40$\deg$ data
(filled squares in Fig.~\ref{fig:qsqdep}).  While there are data at 50$\deg$
(open circles) for all nuclei, the statistical precision is noticeably worse,
and there are much larger corrections for charge symmetric background and
Coulomb distortion (for heavier nuclei).

\begin{figure}[tbh]
\centering
\includegraphics[width=85mm]{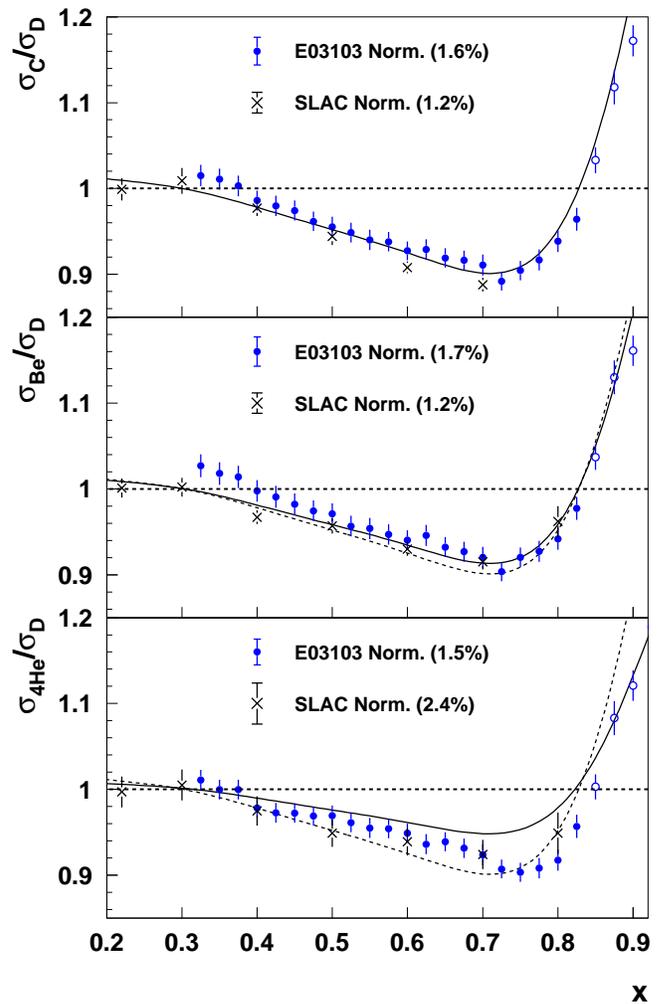}
\caption{(Color online) EMC ratios for $^{12}$C, $^9$Be, and
$^4$He~\cite{note_webdata}, compared to SLAC~\cite{gomez94}.  The $^9$Be
results include a correction for the neutron excess (see text). Closed (open)
circles denote $W^2$ above (below) 2~GeV$^2$.  The solid curve is the
A-dependent fit to the SLAC data, while the dashed curve is the fit to
$^{12}$C. Normalization uncertainties are shown in parentheses for both
measurements.}
\label{fig:c12he4}
\end{figure}

The EMC ratios for $^{12}$C, $^9$Be, and $^4$He are shown in
Fig.~\ref{fig:c12he4} along with results from previous SLAC extractions.  The
$^4$He and $^{12}$C results are in good agreement with the SLAC results, with
much better precision for $^4$He in the new results.  While the agreement for
$^9$Be does not appear to be as good, the two data sets are in excellent
agreement if we use the same isoscalar correction as E139 (see below) and take
into account the normalization uncertainties in the two data sets. In all
cases, the new data extend to higher $x$, although at lower $W^2$ values than
the SLAC ratios. The EMC ratio for $^4$He is comparable to 
$^{12}$C, suggesting that the modification is dependent on the average
nuclear density, which is similar for $^4$He and $^{12}$C, rather than a
function of nuclear mass.

\begin{figure}[tbh]
\centering
\includegraphics[width=85mm]{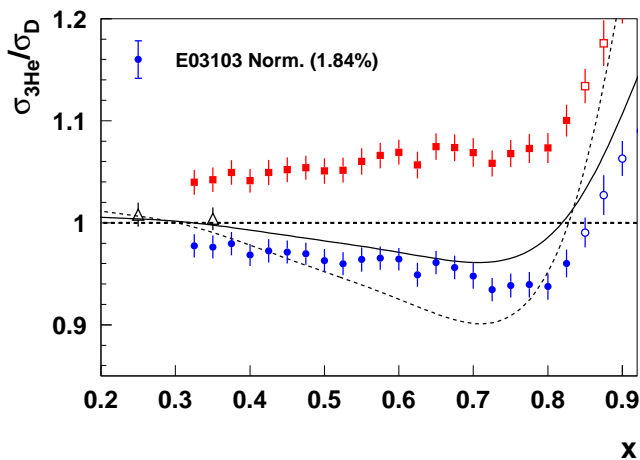}
\caption{(Color online) EMC ratio for $^3$He~\cite{note_webdata}.  The upper
squares are the raw $^3$He/$^2$H ratios, while the bottom circles show the
isoscalar EMC ratio (see text).  The triangles are the HERMES
results~\cite{ackerstaff00} which use a different isoscalar correction.  The
solid (dashed) curves are the SLAC A-dependent fits to Carbon and $^3$He.}
\label{fig:he3}
\end{figure}

Figure~\ref{fig:he3} shows the EMC ratio for $^3$He, with the low-$x$ data
from HERMES.  Note that the HERMES $^3$He data have been renormalized by a
factor of 1.009 based on comparisons of their $^{14}$N EMC effect and the NMC
$^{12}$C result~\cite{ackerstaff00}. We show both the
measured cross section ratio (squares) and the ``isoscalar'' ratio (circles),
where the $^3$He result is corrected for the proton excess.
Previous high-$x$ EMC measurements used a correction based on an extraction
of the $F_{2n}/F_{2p}$ ratio for free nucleons from high $Q^2$ measurements
of $F_{2d}/F_{2p}$.  We use global
fits~\cite{christy07,bosted08} to the free proton and neutron cross
sections evaluated at the kinematics of our measurement and then broadened
using the convolution procedure of Ref.~\cite{arrington09} to yield the 
neutron-to-proton cross section ratio in nuclei.  Using the ``smeared''
proton and neutron cross section ratios more accurately reflects the
correction that should be applied to the nuclear ratios, and in the end,
yields a significantly smaller correction at large $x$, where the uncertainty
in the neutron structure function is largest.  

While applying the isoscalar correction to the $^3$He data, using the
smeared $F_{2n}/F_{2p}$ ratio, yields a more reliable result, there is
still some model dependence to this correction due to the 
uncertainty in our knowledge of the neutron structure function.
Ref.~\cite{arrington09} demonstrated that much of the inconsistency between
different extractions of the neutron structure function comes from 
comparing fixed-$Q^2$ calculation to data with varying
$Q^2$ values, rather than from the underlying assumptions of nuclear effects
in the deuteron.  Nuclear effects beyond what is included in
Ref.~\cite{arrington09}, such as the off-shell contribution $\delta^{(off)}$
of Ref.~\cite{melnitchouk94}, yield a 1--2\% decrease to the proton's
contribution to the deuteron thus increasing the extracted $F_{2n}/F_{2p}$
ratio by 0.01--0.02.  This yields a slightly reduce correction for $^3$He
which would raise the isoscalar EMC ratio for $^3$He by 0.3--0.6\% at
our kinematics.

The observed nuclear effects are clearly smaller for $^3$He than for $^4$He
and $^{12}$C.  This is again consistent with models where the EMC effect
scales with the average density, as the average density for $^3$He is roughly
half that of the $^{12}$C.  However, the results of $^9$Be are not consistent
with the simple density-dependent fits. The observed EMC effect in $^9$Be is
essentially identical to what is seen in $^{12}$C, even though the density of
$^9$Be is much lower.  This suggests that both the simple mass- or
density-scaling models break down for light nuclei.

\begin{figure}[tbh]
\centering
\includegraphics[width=85mm]{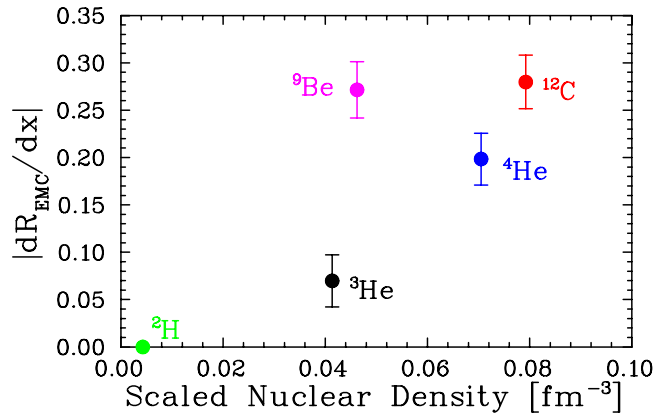}
\caption{(Color online) The circles show the slope of the isoscalar EMC ratio
for $0.35<x<0.7$ as a function of nuclear density.  Error bars include
statistical and systematic uncertainties.}
\label{fig:adep}
\end{figure}

One can examine the nuclear dependence based on the size of the EMC ratio at a
fixed $x$ value, but the normalization uncertainties become a significant
limiting factor.  If we assume that the shape of the EMC effect is universal,
and only the magnitude varies with target nucleus, we can
compare light nuclei by taking the $x$ dependence of the ratio in the linear
region, $0.35 < x < 0.7$, using the slope as a measure of the relative size
of the EMC effect that is largely unaffected by the normalization.  The slopes
are shown for light nuclei in Fig.~\ref{fig:adep} as a function of average
nuclear density.  The average density is calculated from the \textit{ab initio}
GFMC calculation of the spatial distributions~\cite{pieper01}.  Because
we expect that it is the presence of the other $(A-1)$ nucleons that yields
the modification to the nuclear structure function, we choose to
scale down this density by a factor of $(A-1)/A$, to remove the struck
nucleon's contribution to the average density.  The EMC effect for $^3$He is
roughly one third of the effect in $^4$He, in contrast to the A-dependent fit
to the SLAC data~\cite{gomez94}, while the large EMC effect in $^9$Be
contradicts a simple density-dependent effect.

One explanation for the anomalous behavior of $^9$Be is that it can be
described as a pair of tightly bound alpha particles plus one additional
neutron~\cite{arai96}.  While most of the nucleons are in a dense environment,
similar to $^4$He, the \textit{average} density is much lower, as the
alphas (and additional neutron) `orbit' in a larger volume.  This
suggests that it is the \textit{local} density that drives the modification. 
The strong clustering of nucleons in $^9$Be leads to a special case where
the average density does not reflect the local environment of the bulk of the
protons and neutrons.  

Another possibility is that the $x$ dependence of the EMC effect is
different enough in these light nuclei that we cannot use the falloff with
$x$ as an exact measure of the relative size of the EMC effect.  This too
suggests that the EMC effect is sensitive to the details of the nuclear
structure, which would require further theoretical examination.  At the
moment, there are almost no calculations for light nuclei that include detailed
nuclear structure.

In conclusion, we have measured the nuclear dependence of the structure
functions for a series of light nuclei.  This data set provides significantly
improved data on $^4$He and the first valence-region measurement on $^3$He,
as well as extending the measurements to higher $x$ for other light nuclei.
This will allow for more detailed comparison with calculations that include
binding and Fermi motion, providing a more reliable baseline at low $x$, where
these effects are still important, but may not fully explain the observed
nuclear dependence.

These data also provide model independent information on the scaling of the
nuclear effects.  Under the assumption that the shape of the EMC effect is the
same for all nuclei, the large difference between $^3$He and $^4$He rules out
previous A-dependent fits, while the EMC effect in $^9$Be is inconsistent
with models where the effect scales with average density.
The results are consistent with the idea that the effect scales with the
\textit{local} environment of the nucleons, or require that the $x$-dependence
of the effect changes in very light nuclei.

\begin{acknowledgments}

This work was supported in part by the NSF and DOE, including DOE contract
DE-AC02-06CH11357, DOE contract DE-AC05-06OR23177 under which JSA, LLC
operates JLab, and the South African National Research Foundation.

\end{acknowledgments}

\bibliography{prl_e03103}

\end{document}